%% file: main.tex
\pdfoutput=1

\documentclass[11pt]{article}

\usepackage[final]{acl}

\usepackage{times}
\usepackage{latexsym}

\usepackage{enumitem}
\usepackage{booktabs}
\usepackage{xcolor}
\usepackage{soul}
\newcommand{\highlight}[2]{\sethlcolor{#1}\hl{#2}}
\definecolor{trained}{HTML}{DAE8FC}
\definecolor{pretrained}{HTML}{E1D5E7}
\definecolor{nonparam}{HTML}{FFE6CC}

\usepackage{amssymb}
\newlist{checklist}{itemize}{1}
\setlist[checklist]{label=$\square$, itemsep=0pt, parsep=0pt, topsep=0pt, partopsep=0pt}

\usepackage[T1]{fontenc}

\usepackage[utf8]{inputenc}

\usepackage{microtype}

\usepackage{inconsolata}

\usepackage{graphicx}

\usepackage{subcaption}

\title{kNN Retrieval for Simple and Effective Zero-Shot Multi-speaker Text-to-Speech}

\author{
 \textbf{Karl El Hajal\textsuperscript{1,2}},
 \textbf{Ajinkya Kulkarni\textsuperscript{1}},
 \textbf{Enno Hermann\textsuperscript{1}},
 \textbf{Mathew Magimai.-Doss\textsuperscript{1}}
\\
\\
 \textsuperscript{1}Idiap Research Institute, CH-1920 Martigny, Switzerland \\
 \textsuperscript{2}EPFL, \'Ecole polytechnique f\'ed\'erale de Lausanne, CH-1015 Lausanne, Switzerland
 \\
\\
 \texttt{
    \{karl.elhajal,enno.hermann,ajinkya.kulkarni,mathew\}@idiap.ch
 }
}

\begin{document}
\maketitle
\begin{abstract}
While recent zero-shot multi-speaker text-to-speech (TTS) models achieve impressive results, they typically rely on extensive transcribed speech datasets from numerous speakers and intricate training pipelines. Meanwhile, self-supervised learning (SSL) speech features have emerged as effective intermediate representations for TTS. Further, SSL features from different speakers that are linearly close share phonetic information while maintaining individual speaker identity. In this study, we introduce kNN-TTS, a simple and effective framework for zero-shot multi-speaker TTS using retrieval methods which leverage the linear relationships between SSL features. Objective and subjective evaluations show that our models, trained on transcribed speech from a single speaker only, achieve performance comparable to state-of-the-art models that are trained on significantly larger training datasets. The low training data requirements mean that kNN-TTS is well suited for the development of multi-speaker TTS systems for low-resource domains and languages. We also introduce an interpolation parameter which enables fine-grained voice morphing.
Demo samples are available at \href{https://idiap.github.io/knn-tts/}{https://idiap.github.io/knn-tts}.
\end{abstract}

\section{Introduction}

Neural text-to-speech (TTS) synthesis has advanced significantly in recent years, achieving a level of naturalness comparable to human speech. and allowing for an increasingly expressive range of outputs \cite{tan2021survey}.
Neural TTS systems can be categorized into two-stage and single-stage pipelines. Two-stage models convert text or phonemic features into acoustic features and then use a vocoder to generate waveforms. These models can suffer from error propagation and limitations due to their dependence on low-level features like mel-spectrograms \cite{glowtts, tacotron2}.
Single-stage models aim to address these issues by streamlining this process into an end-to-end framework \cite{vits, pmlr-v162-casanova22a}, but they may face oversmoothing, mispronunciations, and reduced flexibility due to the lack of explicit linguistic information and entangled latent representations \cite{lee2022hierspeech, choi2023nansy}.
Recent research combines the strengths of both approaches by using self-supervised learning (SSL) speech representations as intermediate elements in two-stage models \cite{wavthruvec, parrottts, comparative_study_ssl_tts}.
These representations help improve word error rates, pronunciation of out-of-vocabulary words \cite{wavthruvec}, and robustness to noise \cite{zhu2023rep2wav}.

In practice, end-user applications may need to synthesize speech in the voices of multiple speakers. Collecting high quality speech data and building a TTS model for each target voice is a challenging problem. As a result, there has been a growing interest in zero-shot multi-speaker TTS systems which can synthesize speech in an unseen speaker's voice based on short reference samples. State-of-the-art models such as XTTS \cite{casanova2024xtts} and HierSpeech++~\cite{lee2023hierspeechbridginggapsemantic} demonstrate impressive quality and similarity to unseen speakers. To produce varied voices, these models condition the output on style embeddings, which are extracted from a reference audio sample via a speaker encoder. However, these models require end-to-end training on thousands of hours of transcribed audio data from a large number of speakers to generalize effectively. 

Simultaneously, kNN-VC \cite{baas23_interspeech} has emerged as a promising any-to-any voice conversion method, leveraging SSL features for zero-shot conversion. It uses a kNN algorithm to match frames from the source speaker with the target speaker's representations, adjusting the speaker identity while preserving speech content. This approach is similar to retrieval-augmented generation (RAG) techniques used in deep generative models such as language models \cite{khandelwal2020knnlm, khandelwal2021knnmt} and image generators \cite{chen2022reimagen}. These methods have been effectively used in these fields to enhance accuracy and reliability, as well as to enable style transfer by steering model outputs to mirror characteristics of a retrieval database \cite{borgeaud2022improving, chen2022reimagen}.

In this work, we investigate whether retrieval-based methods can be similarly applied to TTS for style-transfer, to achieve effective zero-shot multi-speaker capabilities.
Additionally, we explore whether these methods can reduce data requirements for the development of a robust zero-shot multi-speaker TTS system.
This paper's key contributions can be summarized as follows:

\begin{itemize}[leftmargin=10pt]
\itemsep0em 
    \item We propose kNN-TTS, a novel framework for multi-speaker zero-shot TTS which leverages retrieval methods to modify target voices, diverging from the conventional approach of using speaker embeddings.

    \item By exploiting linear relationships in SSL features, our framework alleviates the need for multi-speaker transcribed data during training.

    \item We introduce a novel linear interpolation parameter allowing for fine-grained control over the influence of the target style on the output, which offers voice morphing capabilities.

    \item We validate the method using two different lightweight models trained solely on transcribed speech from one speaker and demonstrate competitive performance with state-of-the-art models trained on much larger datasets.    
\end{itemize}

Code, models, and demo samples are publicly available at \href{https://idiap.github.io/knn-tts}{https://idiap.github.io/knn-tts}.

\section{Proposed Approach}
\subsection{Framework}
The kNN-TTS framework, illustrated in Fig.~\ref{fig:model_architecture}, begins with a Text-to-SSL model that generates source speaker features from text input. A kNN retrieval algorithm then matches these generated features to units in a target speaker's unit database, which contains features extracted from the target speaker's recordings using a pre-trained SSL encoder. The selected target speaker features are linearly interpolated with the source speaker features to obtain the converted features. Finally, a pre-trained vocoder decodes the converted features back into a speech waveform.

\textbf{SSL encoder:} For this framework, we need an intermediate audio representation that meets the following criteria: (1) it should encompass both linguistic and speaker-specific information; (2) features that are linearly close should exhibit similar phonetic properties while preserving speaker identity; and (3) it should be possible to decode the features back to waveform. Recent works show that SSL models encode speech into representations that meet these criteria \cite{dunbar_2022}. Preliminary experiments indicate that spectral features are ineffective in this context (Appendix \ref{sec:appendix_spectral_features}).

\textbf{Text-to-SSL:} We train a Text-to-SSL model that generates corresponding SSL features from a given text input. Notably, this is the only component of our framework that requires audio data paired with text transcriptions for training. It is possible to train this model on the speech of a single speaker. 

\textbf{kNN Retrieval:} To synthesize speech in a target speaker's voice, units (or frames) from the target speaker unit database are selected to replace corresponding frames from the source speaker features. The selection is done by comparing source and target frames using a linear distance metric. This results in selected target speaker features that maintain the phonetic information while replacing the voice attributes with those of the target speaker.
\input{figures/framework_overview}

The source and target speaker features are then linearly interpolated to obtain the converted features~\cite{khandelwal2020knnlm}. A variable parameter $\lambda$ modifies the degree of influence the target features have on the output, enabling voice morphing by blending the source and target styles. 
\begin{equation}
{y}_{\mathrm{converted}} = \lambda  \  {y}_{\mathrm{selected}} + (1-\lambda) \ {y}_{\mathrm{source}} 
\end{equation}

\textbf{Vocoder:} We employ a vocoder capable of decoding the SSL features back into a waveform. To ensure robust generalization, the vocoder should be pre-trained on a large and diverse dataset to maintain high-quality waveform reconstruction across different speakers and contexts.

\subsection{Implementation}

\textbf{SSL encoder:} We employ a pre-trained WavLM-Large encoder from \cite{wavlm}.
It is specifically selected due to its effective audio reconstruction capabilities, obtained through training on masked speech denoising and prediction tasks \cite{wang23_ssw}.
We use the features from the model's 6th layer which encapsulate both phonetic and speaker characteristics~\cite{baas23_interspeech,wang23_ssw}.
These representations are pre-extracted and cached prior to training or inference, eliminating the need to load WavLM during either process, assuming the target speaker is known.

\textbf{Text-to-SSL:} We evaluate two Text-to-SSL implementations: GlowTTS \cite{glowtts} and GradTTS \cite{pmlr-v139-popov21a-gradtts}. GlowTTS employs a non-autoregressive architecture with a transformer-based text encoder, a duration predictor, and a flow-based decoder \cite{kingma2018glow}. GradTTS follows a similar architecture but uses a diffusion-based decoder \cite{song2021scorebased}.
We maintain each model's default configurations and cost functions for training. We adjust only their output dimension to 1024 channels to align with WavLM-Large features instead of mel-spectrograms. For the GradTTS diffusion decoder, we use 100 iterations for synthesis. Both models are trained on the LJSpeech dataset \cite{ljspeech17}, which comprises 24~hours of single-speaker English speech. GlowTTS is trained for 650k steps, and GradTTS for 2M steps.

\textbf{kNN Retrieval:} For each source frame, we compute its cosine distance with every target speaker frame within the unit database. We then select the $k$ closest units, and average them with uniform weighting.
Similar to \citet{baas23_interspeech}, we use $k = 4$ which was determined to be suitable across different amounts of target audio.

\textbf{Vocoder:} We use a pre-trained HiFi-GAN V1 \cite{hifigan} model trained to reconstruct 16kHz waveforms from WavLM-Large layer 6 features. The model checkpoint, sourced from \citet{baas23_interspeech}, was trained using their pre-matched paradigm on the LibriSpeech train-clean-100 set, consisting of 100 hours of clean English speech from 251 speakers \cite{librispeech}.

\section{Experimental Setup}
\subsection{Baselines}

We benchmark our models against leading open-source zero-shot multi-speaker TTS systems. \textbf{YourTTS} \cite{pmlr-v162-casanova22a} is trained on 529 hours of multilingual transcribed data from over 1000 speakers. \textbf{XTTS} \cite{casanova2024xtts} uses 27,282 hours of transcribed speech data across 16 languages. \textbf{HierSpeech++} \cite{lee2023hierspeechbridginggapsemantic} is trained on 2796 hours of transcribed English and Korean speech, encompassing 7299 speaker. 
These models are trained end-to-end,
and employ various speaker encoders to convert a reference utterance into a style embedding for zero-shot multi-speaker synthesis. We use the default checkpoints and configurations provided by the authors for each baseline model\footnote{ \href{https://github.com/idiap/coqui-ai-TTS}{https://github.com/idiap/coqui-ai-TTS}}~\footnote{\href{https://github.com/sh-lee-prml/HierSpeechpp}{https://github.com/sh-lee-prml/HierSpeechpp}}. Further details about the baselines can be found in Table~\ref{tab:model_comp} and Appendix~\ref{sec:appendix_baselines}.

\subsection{Evaluation}
\label{sec:eval}

\input{tables/results_multispeaker}

For zero-shot multi-speaker synthesis comparisons, we use LibriSpeech test-clean for target speaker reference utterances. It includes speech of varied quality from 20 male and 20 female speakers, with 8~mins of speech per speaker. For each model, we synthesize 100 English sentences per speaker, selecting the sentences randomly from FLoRes+ \cite{costa2022no}, as per the XTTS protocol. Tests are performed with $\lambda=1$.
For baseline models, we obtain a speaker embedding by averaging style embeddings across all reference utterances of each target speaker, ensuring a fair comparison.

\textbf{Objective analysis:} we evaluate each model's performance in terms of naturalness using UTMOS \cite{saeki22c_interspeech}, intelligibility using the word error rate (WER) and phoneme error rate (PER) computed with the Whisper-Large v3 model \cite{whisper}, and speaker similarity using speaker encoder cosine similarity (SECS) with ECAPA2 \cite{thienpondt2023ecapa2}.

\textbf{Subjective evaluation:} we conduct a listening test to assess naturalness and similarity mean opinion scores (N-MOS and S-MOS). We randomly select utterances from 10 male and 10~female target speakers from LibriSpeech test-clean, choosing 3 synthesized sentences per speaker, totaling 60~utterances per model. Each is rated by 10 raters on naturalness and similarity to a ground-truth recording, with scores ranging from 1 to 5 in 0.5 increments. We use Amazon Mechanical Turk, with raters required to be native English speakers based in the United States, having a HIT acceptance rate above 98\% and more than 100 approved HITs.~Further details are presented in Appendix~\ref{sec:appendix_listening_test}.

\textbf{Model efficiency:} we compare models on parameter count, peak GPU memory usage during test sample synthesis, and real-time factor (RTF), tested on an NVIDIA RTX3090 GPU.

\textbf{Voice Morphing:} we perform an experiment using the interpolation parameter, computing the SECS of the model's output with the target speaker's ground truth data for various values of $\lambda$.

\input{figures/similarity_matrix}
\vspace{-10pt}
\section{Results and analysis}

Results are presented in Table~\ref{tab:model_comp}. Objective metrics reveal that the kNN-TTS models demonstrate the best speaker similarity, XTTS excels in intelligibility, and HierSpeech++ achieves the highest naturalness. In the listening test, HierSpeech++ was rated highest for naturalness and similarity, while the kNN-TTS models and XTTS performed similarly. These models' results fall within each other's confidence intervals, suggesting comparable performance.~Regarding model efficiency, kNN-TTS models have the fewest parameters and lowest memory usage among the top performers. GlowkNN-TTS uses $3\times$ less memory than HierSpeech++ with similar speed.~GradkNN-TTS's memory usage and RTF are higher due to the 100 iterations used in the diffusion decoder. Further, the kNN-TTS models are trained on 100$\times$ less transcribed data than HierSpeech++ and 1000$\times$ less data than XTTS.

Figure~\ref{fig:sim_matrix} illustrates the results of the voice morphing experiment. We can observe that the similarity of the outputs to the target speaker gradually increases as $\lambda$ rises, demonstrating the ability to finely blend source and target styles and suggests the potential to combine multiple target styles.

\section{Discussion and conclusions}

State-of-the-art zero-shot multi-speaker TTS models rely on large datasets of transcribed speech from thousands of speakers for training. In this paper, we demonstrated that by leveraging retrieval methods and SSL features, we can develop a simple and lightweight TTS system that achieves a comparable level of naturalness and similarity to leading approaches while being trained on transcribed data from only a single speaker.
We further showed that fine-grained voice morphing can be achieved using an interpolation parameter. This indicates that this technique, which is originally inspired from other domains such as language modeling~\cite{khandelwal2020knnlm} and machine translation \cite{khandelwal2021knnmt}, can be applied in the context of TTS.

The simplicity of the training process is a main advantage of our approach: only the Text-to-SSL model needs training, and it can be trained on transcribed data from one speaker. In conjunction with the kNN approach's cross-lingual capability \cite{knnvc_followup}, this is particularly appealing for extending the model to new  languages with less resources, a direction open for future work.

We also showed that the framework can be implemented using different Text-to-SSL architectures, allowing for model swapping to leverage different benefits. Our implementations notably demonstrated efficiency in terms of parameters, memory usage, and runtime speed in the case of GlowkNN-TTS, even without optimizing the retrieval process. 

\clearpage

\section*{Limitations}

\subsection*{Reference Data Requirements}
While our approach offers simplicity in training and is more lightweight, it requires more reference audio compared to other methods. We conduct ablation studies to evaluate the models' outputs with varying amounts of reference utterances. Figure~\ref{fig:ljspeech_plot} compares outputs using retrieval from different amounts of LJSpeech data. We find that approximately 30 seconds of reference utterances are needed to achieve suitable intelligibility, while naturalness improves up to 5 minutes, surpassing the model outputs without retrieval. Figure~\ref{fig:libri4077_plot} compares the kNN-TTS models to the baselines for different amounts of reference utterances from a target speaker. Similarly, about 30~seconds are required for suitable intelligibility, while similarity plateaus at around 1 minute. In contrast, the baselines benefit less from increasing the amount of reference utterances beyond 10 to 30 seconds.
There is therefore a trade-off; our method requires at least 30 seconds of reference audio, whereas competing approaches can function with smaller amounts.

\subsection*{Rhythmic variations}

Typically, different speakers exhibit different pronunciation durations. In our method, the duration aspect is determined by the Text-to-SSL model, and the target voice is modified through frame-by-frame selection, meaning that the duration of each utterance remains unchanged for different speakers. Our future work will explore techniques, such as Urhythmic~\cite{niekerk2023}, to address this limitation.

\subsection*{Training Simplicity and Model Capacity}
In this study, we trained and evaluated Text-to-SSL models on transcribed speech from a single speaker to demonstrate that strong performance can be achieved in a simplified low-resource setting. However, expanding the training data to include multiple speakers and larger datasets can increase the model's output quality and enable it to generate speech with a wider range of expressiveness. Similarly, while we prioritized lightweight models for efficiency, more complex models could improve speech quality at the cost of efficiency. These aspects can be explored further in future work.

\input{figures/combined_plots}
\section*{Ethics Statement}

Zero-shot multi-speaker TTS systems such as the one we describe in this manuscript can offer benefits in accessibility, entertainment and education by enabling the generation of varied expressive synthetic voices from textual input. Our approach's lowered data requirements can unlock these benefits for low-resource domains, while its reduced compute needs ensure sustainability.
However, this technology's accessibility also poses many risks, including voice cloning without consent, impersonation, and the creation of deepfake audio for misinformation and manipulation. We note that compared to other zero-shot methods, our proposed approach, requires more data from the target speaker for sufficient quality, reducing impersonation risks. In our research, we strictly adhere to using only public datasets with appropriate licenses.
To mitigate potential harm, it is important to advance research in watermarking synthetic outputs for traceability and developing methods to differentiate synthetic speech from authentic recordings, thereby reducing risks to individuals and groups. 

\section*{Acknowledgement}
This work was partially supported by the Swiss National Science Foundation grant agreement no. 219726 on ``Pathological Speech Synthesis (PaSS)'' and the Innosuisse flagship grant agreement no. PFFS-21-47 on ``Inclusive Information and Communication
Technologies (IICT)''.

\bibliography{main}

\appendix

\section*{Appendix}
\label{sec:appendix}

\section{Spectral Features}
\label{sec:appendix_spectral_features}

We conducted preliminary experiments to assess the viability of spectral features as intermediate representations within our framework. We use a GlowTTS model and HiFi-GAN vocoder that use mel-spectrograms as feature representations.  Table \ref{tab:melspec} presents the outcomes of replicating the experiment described in Section \ref{sec:eval} using mel-spectrogram features instead of SSL features, comparing them with ground truth samples and GlowkNN-TTS outputs. The objective metrics reveal that the resulting speech is unintelligible and of poor quality, demonstrating that these spectral features are unsuitable for our framework. Indeed, they do not meet the requirement of having phonetic similarity while maintaining individual speaker characteristics when linearly close. This helps highlight the importance of using SSL features in this context, as they possess useful properties that align with our defined criteria.

\input{tables/results_melspec}

\section{Model and Training Details}

Table \ref{tab:model_card} presents the detailed configurations for each model. We trained the models using a single NVIDIA RTX 3090 GPU. For both models, we retained the default parameters from their open-source implementations\footnote{\href{https://github.com/huawei-noah/Speech-Backbones}{https://github.com/huawei-noah/Speech-Backbones}}\footnote{\href{https://github.com/coqui-ai/TTS}{https://github.com/coqui-ai/TTS}}, only adjusting their output channels to $1024$ to match the dimension of WavLM-Large features. We pre-processed all audio data by resampling it to 16 kHz, trimming silences from the beginning and end using a Voice Activity Detector, and normalizing the loudness to -20 dB.

\input{tables/model_card}

\section{Baselines Details}
\label{sec:appendix_baselines}

\textbf{YourTTS} \cite{pmlr-v162-casanova22a} builds on VITS \cite{vits}, adding elements for multilingual training and zero-shot multi-speaker capabilities. It uses the H/ASP speaker encoder \cite{chung2020in}, pre-trained on the VoxCeleb2 dataset \cite{chung18b_interspeech}, to extract a speaker embedding from reference utterances. This embedding conditions the model's duration predictor, flow-based decoder, posterior encoder, and vocoder.

\textbf{XTTS} \cite{casanova2024xtts} features a Vector Quantised-Variational AutoEncoder (VQ-VAE) that encodes mel-spectrograms into discrete codes, a GPT-2 encoder that predicts these audio codes from text tokens, and a HiFi-GAN-based decoder. The GPT-2 encoder is conditioned on speaker information using a Perceiver conditioner, which outputs 32 1024-dimensional embeddings from a mel-spectrogram. The decoder is also conditioned on a speaker embedding extracted using H/ASP.

\textbf{HierSpeech++} \cite{lee2023hierspeechbridginggapsemantic} comprises a text-to-vec module and a hierarchical speech synthesizer. 
The text-to-vec module generates massively multilingual speech (MMS) representations \cite{pratap2024scaling} from text inputs and prosody prompts. The hierarchical speech synthesizer produces a waveform from MMS features and a style prompt. 
Prosody and voice style representations are extracted from reference mel-spectrograms using style encoders comprising 1D convolutional networks, a multi-head self-attention temporal encoder, and a linear projection.

\section{Listening Test}
\label{sec:appendix_listening_test}
To ensure reliable ratings, we implemented the following measures:

\begin{itemize}
    \item Recruited native English speakers from the United States via Mechanical Turk.
    \item Required participants to have >100 approved HITs and a >98\% approval rate.
    \item Compensated raters at \$15/hour (\$0.5 per 2-minute task), exceeding the U.S. minimum wage.
    \item Clearly defined task objectives at the start and alongside each question.
    \item Added a sound check and training samples at the beginning of the test to help the raters adjust to the tasks.
    \item Included attention check samples with specific audio instructions (e.g., "This is an attention check, please select the number 3 to confirm your attention"). Raters were informed about the presence of such checks at the beginning of the listening test.
    \item Filtered out unreliable raters based on attention check performance and ground truth sample ratings.
\end{itemize}

\subsubsection*{Rating Criteria}

\paragraph{Naturalness:} Participants rated audio clips on a scale from 1 (Bad) to 5 (Excellent) with 0.5 increments. The prompt was:

\begin{quote}
\textit{Rate how natural each audio clip sounds on a scale from 1 (Bad) to 5 (Excellent). Excellent indicates completely natural speech, and Bad indicates completely unnatural speech. In this context, Naturalness refers to whether the speech sounds like it's produced by a native English-speaking human.}
\end{quote}
Rating options were:
\begin{checklist}
    \item 5 - Excellent - Completely natural speech
    \item 4.5
    \item 4 - Good - Mostly natural speech
    \item 3.5
    \item 3 - Fair - Equally natural and unnatural speech
    \item 2.5
    \item 2 - Poor - Mostly unnatural speech
    \item 1.5
    \item 1 - Bad - Completely unnatural speech
\end{checklist}

\paragraph{Similarity:} Raters compared each clip to a reference voice, using the same scale. The prompt was:

\begin{quote}
\textit{Compare each audio clip with the reference voice. Rate whether you feel they are spoken by the same speaker on a scale from 1 (Bad) to 5 (Excellent). Excellent indicates exactly the same speaker, and Bad indicates completely different speakers.}
\end{quote}
Rating options were:
\begin{checklist}
    \item 5 - Excellent - Identical to reference speaker
    \item 4.5
    \item 4 - Good - Mostly similar to reference speaker
    \item 3.5
    \item 3 - Fair - Somewhat different from reference speaker
    \item 2.5
    \item 2 - Poor - Mostly unlike reference speaker
    \item 1.5
    \item 1 - Bad - Completely different from reference speaker
\end{checklist}

\end{document}

%% file: figures/framework_overview.tex
\begin{figure}
    \centering
    \includegraphics[width = \columnwidth]{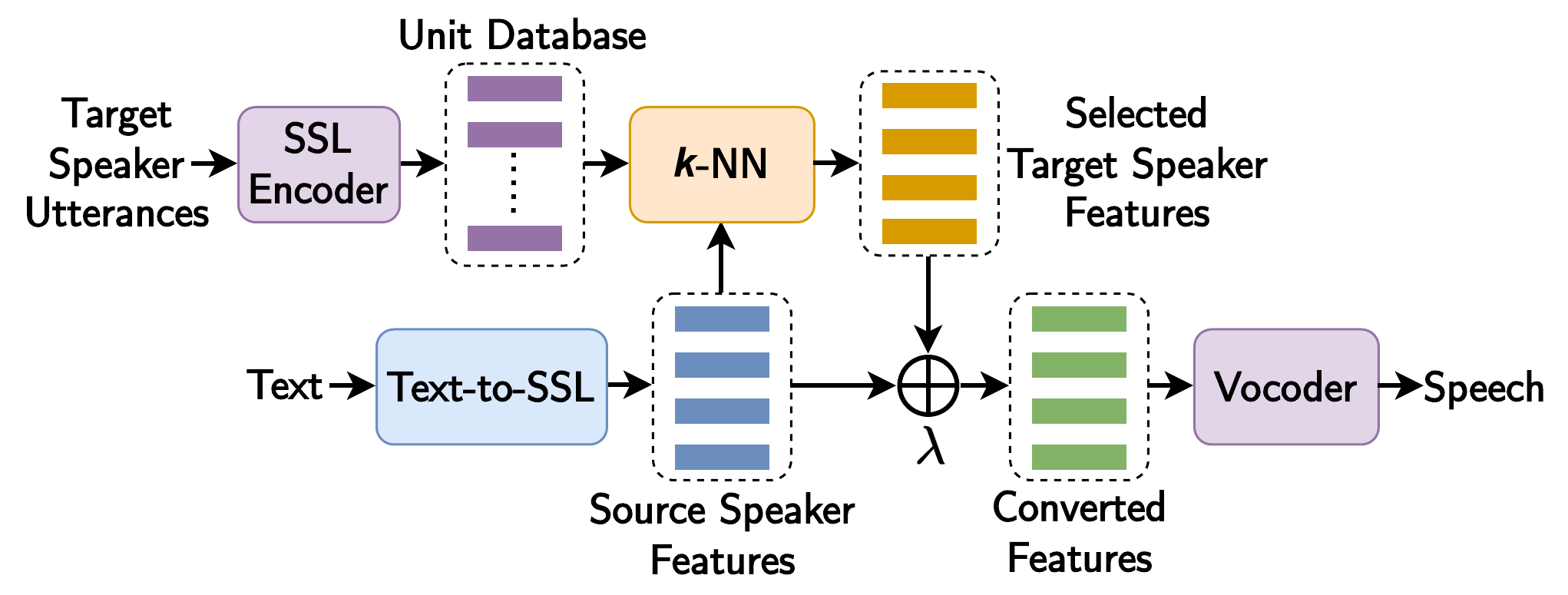}
    \caption{kNN-TTS framework overview.~Only the Text-to-SSL model is \highlight{trained}{trained} on transcribed audio. The SSL encoder, vocoder are \highlight{pretrained}{pre-trained} on untranscribed multi-speaker data, and the kNN algorithm is \highlight{nonparam}{non-parametric}.
    }
    \vspace{-10pt}
    \label{fig:model_architecture}
\end{figure}

%% file: tables/results_multispeaker.tex
\begin{table*}[htbp]
\centering
\caption{Zero-shot multi-speaker TTS results. Training data specifically refers to transcribed data. Evaluation scores are reported with 95\% confidence intervals, and the best scores for each metric are highlighted in bold. 
}
\resizebox{\textwidth}{!}{
\begin{tabular}{|l|cccc|cccc|cc|}
\hline
 & \textbf{\#Params} & \textbf{Training Data} & \textbf{Memory} & \textbf{RTF} & \textbf{WER} & \textbf{PER} &\textbf{UTMOS} & \textbf{SECS} & \textbf{N-MOS} & \textbf{S-MOS} \\
 \textbf{Model} & \textbf{(M)} & \textbf{(Hours)} & \textbf{(GB)} &  & $\downarrow$ & $\downarrow$ & $\uparrow$ & $\uparrow$ & $\uparrow$ & $\uparrow$
\\ \hline

Ground Truth  & n/a & n/a & n/a & n/a & 2.91 $\pm$ 0.31 & 0.92 $\pm$ 0.15 & 4.09 $\pm$ 0.01 &  0.87 $\pm$ 0.003 & 4.21 $\pm$ 0.06 & 4.12 $\pm$ 0.06  \\ \hline

\textit{\textbf{Baselines:}} &&&&&&&&&&\\
YourTTS       & 85.5 & 529 & 0.56 & 0.71 & 6.09 $\pm$ 0.32 & 2.24 $\pm$ 0.12 & 3.65 $\pm$ 0.01 &  0.54 $\pm$ 0.003 & 3.87 $\pm$ 0.08 & 3.86 $\pm$ 0.09 \\
XTTS          & 482  & 27,282 & 2.15 & 1.64 & \textbf{2.76 $\pm$ 0.21} & 0.84 $\pm$ 0.09 & 4.07 $\pm$ 0.01 &  0.40 $\pm$ 0.003  & 4.11 $\pm$ 0.06 & 3.93 $\pm$ 0.08\\
HierSpeech++  & 63  & 2,796 & 1.29 & \textbf{0.18}  & 3.36 $\pm$ 0.23 & \textbf{0.78 $\pm$ 0.06} & \textbf{4.44 $\pm$ 0.01} &  0.67 $\pm$ 0.003 & \textbf{4.15 $\pm$ 0.06} & \textbf{4.01 $\pm$ 0.08} \\ \hline
\textit{\textbf{Proposed:}} &&&&&&&&&&\\

GlowkNN-TTS   & 51.5 & \textbf{24} & \textbf{0.45} & 0.24 & 3.71 $\pm$ 0.24 & 0.98 $\pm$ 0.07 & 4.02 $\pm$ 0.01 &  \textbf{0.72 $\pm$ 0.002} & 4.07 $\pm$ 0.07 & 3.93 $\pm$ 0.08 \\
GradkNN-TTS   & \textbf{31.5} & \textbf{24} & 0.91 & 2.41 & 4.32 $\pm$ 0.25 & 1.44 $\pm$ 0.09 & 4.16 $\pm$ 0.01 &  0.71 $\pm$ 0.003 & 4.10 $\pm$ 0.07 & 3.91 $\pm$ 0.08 \\ \hline
\end{tabular}
}
\label{tab:model_comp}
\end{table*}

%% file: figures/similarity_matrix.tex
\begin{figure}[h]
    \centering
    \includegraphics[width = .6\columnwidth]{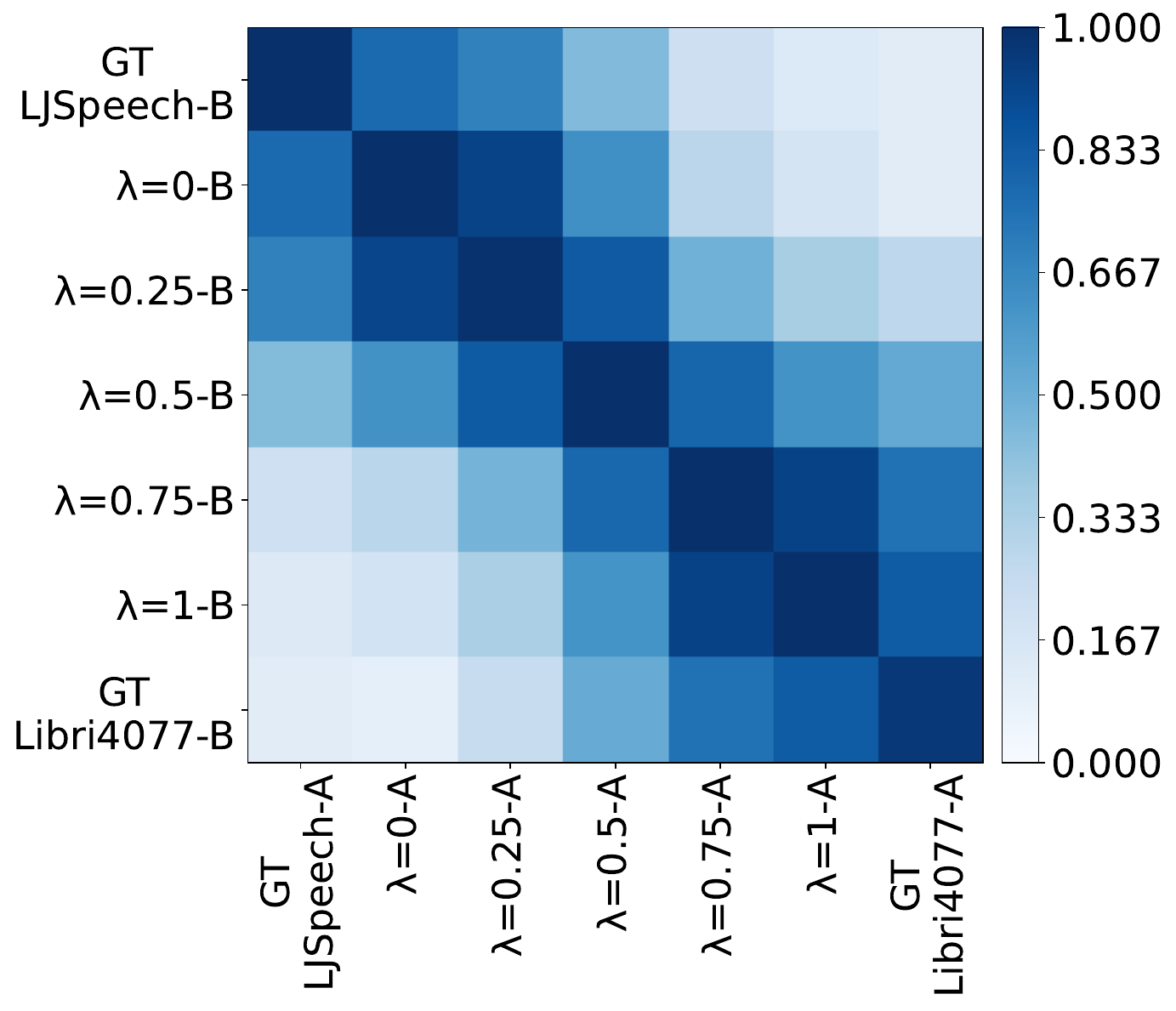}
    \caption{Speaker similarity matrix comparing SECS values for ground truth (GT) LJSpeech samples, LibriSpeech Speaker 4077 (Libri4077) recordings, and GlowkNN-TTS outputs with kNN retrieval from Libri4077 data for various $\lambda$ values. Samples in each case are split in half into sets $A$ and $B$ and compared.}
    \label{fig:sim_matrix}
\end{figure}

%% file: figures/combined_plots.tex
\begin{figure}[t]
  \centering
  \begin{subfigure}{.49\columnwidth}
    \includegraphics[width=\columnwidth]{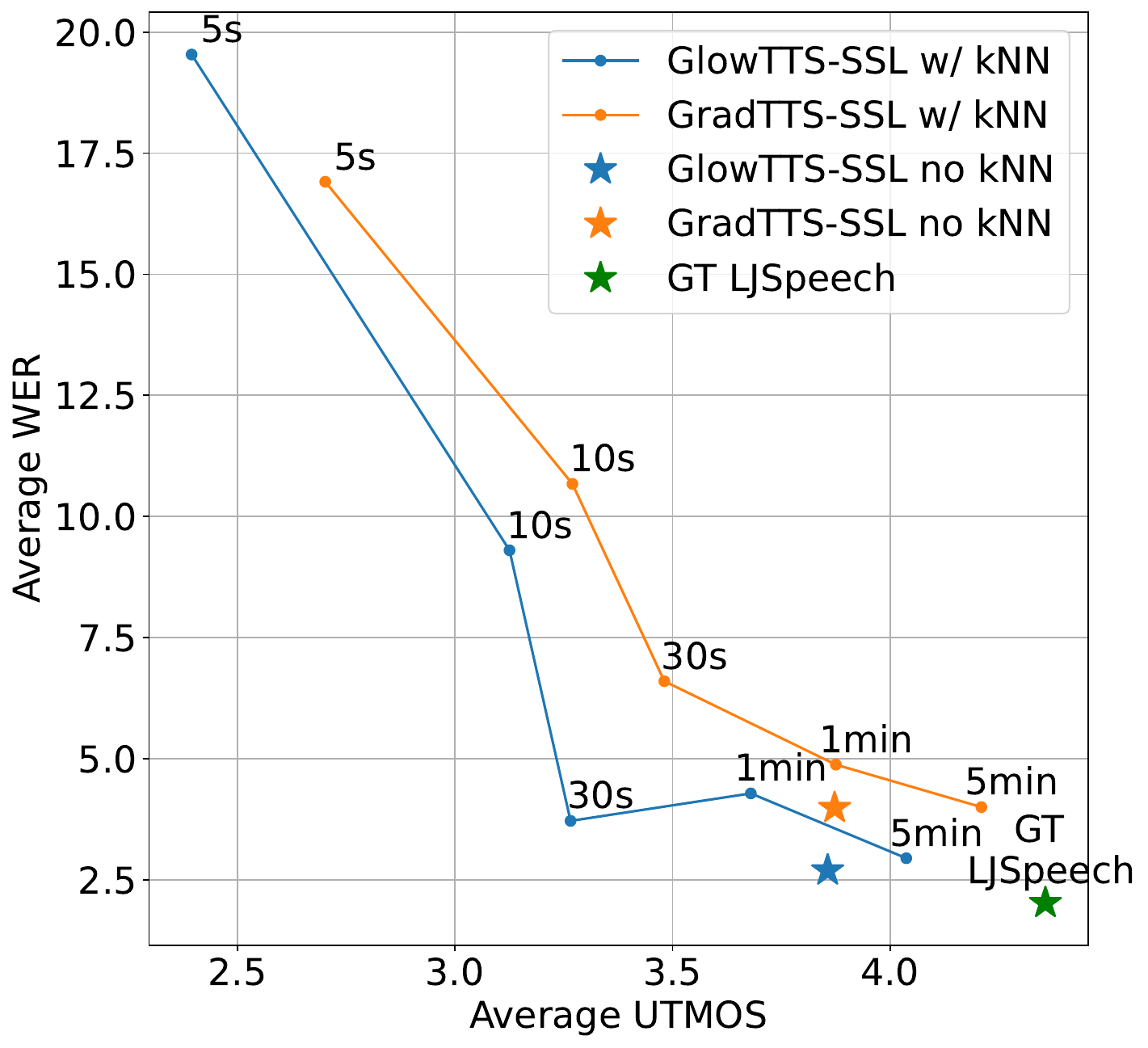}
    \caption{}
    \label{fig:ljspeech_plot}
  \end{subfigure}
   \begin{subfigure}{.48\columnwidth}
    \includegraphics[width=\columnwidth]{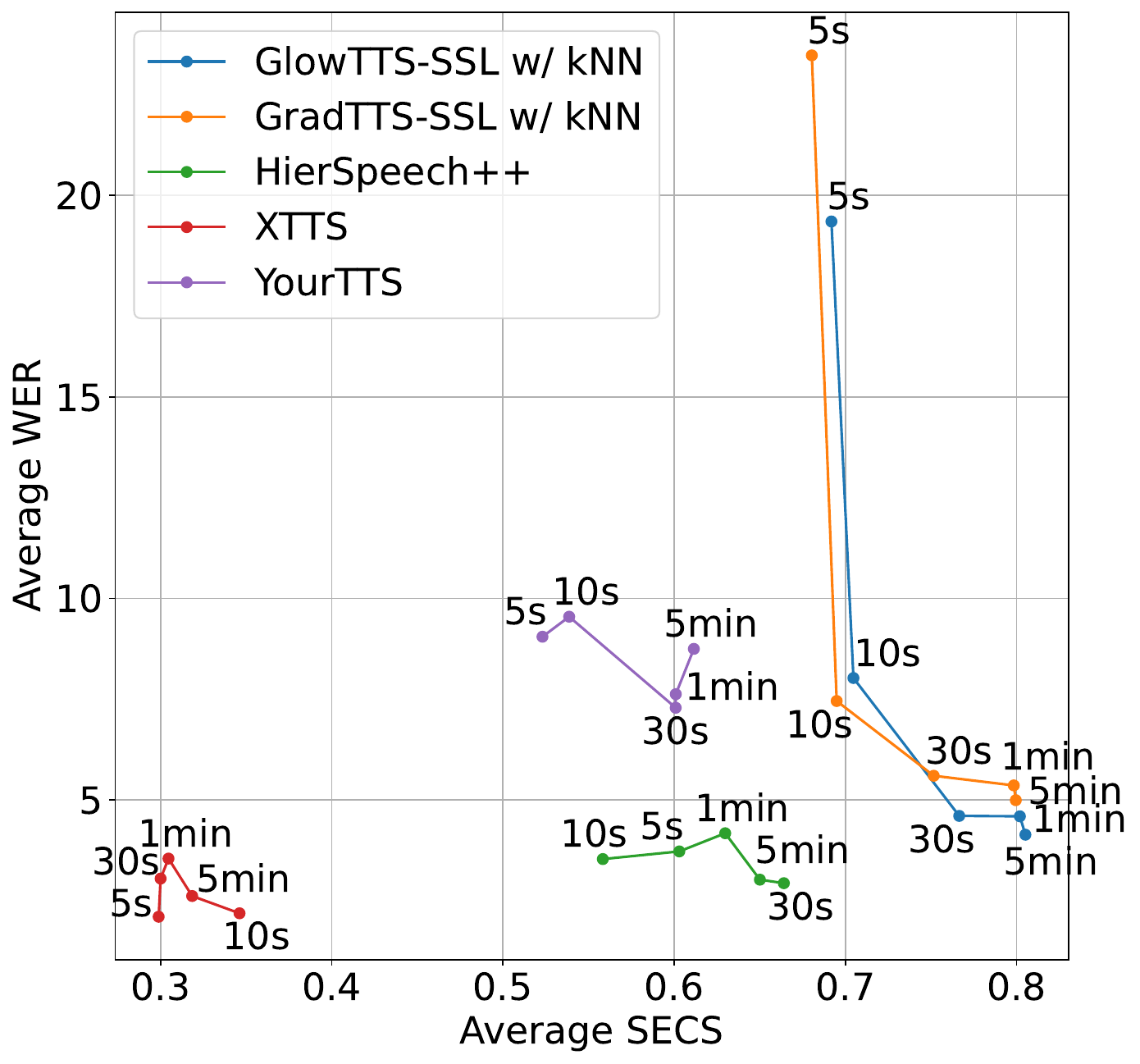}
    \caption{}
    \label{fig:libri4077_plot}
  \end{subfigure}
  
  \caption{
  (a) Mean UTMOS ($\uparrow$) and WER ($\downarrow$) for kNN-TTS outputs using different amounts of LJSpeech reference utterances. (b) Mean SECS ($\uparrow$) and WER ($\downarrow$) for kNN-TTS and baseline outputs using different amounts of LibriSpeech Speaker 4077 reference utterances.}
  \label{fig:plots}
\end{figure}

%% file: tables/results_melspec.tex
\begin{table}[htbp]
\caption{Objective metrics comparing the Ground Truth and GlowkNN-TTS model to the experiment using mel-spectrogram features as intermediate representations (MelSpec).}
\resizebox{\columnwidth}{!}{
\begin{tabular}{|l|cccc|}
\hline
\textbf{Model} & \textbf{WER} ($\downarrow$) & \textbf{PER} ($\downarrow$) &\textbf{UTMOS} ($\uparrow$) & \textbf{SECS} ($\uparrow$) \\
 \hline
Ground Truth & 2.91 $\pm$ 0.3 & 0.92 $\pm$ 0.2 & 4.09 $\pm$ 0.01 &  0.87 $\pm$ 0.003 \\
GlowkNN-TTS & 3.71 $\pm$ 0.2 & 0.98 $\pm$ 0.07 & 4.02 $\pm$ 0.01 &  0.72 $\pm$ 0.002 \\
MelSpec & 109 $\pm$ 5 & 79 $\pm$ 5 & 1.27 $\pm$ 0.001 &  0.15 $\pm$ 0.004\\
\hline
\end{tabular}
}
\label{tab:melspec}
\end{table}

%% file: tables/model_card.tex
\begin{table}
\centering
\caption{Detailed configurations for the GlowkNN-TTS and GradkNN-TTS models presented in this paper.}
\label{tab:model_card}

\resizebox{\columnwidth}{!}{
\begin{tabular}{lcc} 
\toprule
\textbf{Config}               & GlowkNN-TTS        & GradkNN-TTS   \\ 
\hline
Optimiser                     & RAdam              & Adam                  \\
Betas                         & [$0.9, 0.998$]     & n/a                \\
Learning rate                 & $1e^{-3}$          & $1e^{-4}$                \\
Scheduler                     & Noam               & n/a             \\
Batch Size                    & 32                 & 16                      \\
Mixed-precision               & 16bit              & 16bit                  \\ 
Steps                         & 650k               & 2M                      \\
\midrule
\#Parameters                    & 51.5M              & 31.5M                      \\
\textbf{\textit{Encoder}}    &&\\
Hidden Channels               & 192               & 192                      \\
Kernel Size                     &  3 &  3                                     \\
Dropout                         & 0.1   &  0.1 \\
Layers                          & 6     &  6 \\
Heads                           & 2     &  2 \\
FFN Channels                    & 768   &  768 \\
Duration Predictor Channels & 256 & 256 \\
\textbf{\textit{Decoder}}    &&\\
Hidden Channels               & 192                & 64                       \\
Output Channels               & 1024               & 1024                      \\
Dropout                       & 0.05 & n/a \\
Flow Blocks                   & 12 & n/a \\
Kernel Size                   & 5   & n/a \\
$\beta_0$, $\beta_1$          & n/a   & 0.05, 20 \\
\bottomrule
\end{tabular}
}
\end{table}